\documentclass[aps,prl,reprint,superscriptaddress,letterpaper,english,longbibliography,nobibnotes]{revtex4-2}
\usepackage[T1]{fontenc}
\usepackage[utf8]{inputenc}
\DeclareUnicodeCharacter{00A0}{~}
\DeclareUnicodeCharacter{2009}{\,}
\DeclareUnicodeCharacter{2010}{-}
\DeclareUnicodeCharacter{2013}{--}
\DeclareUnicodeCharacter{2019}{'}
\DeclareUnicodeCharacter{201C}{``}
\DeclareUnicodeCharacter{201D}{''}
\DeclareUnicodeCharacter{2212}{\ensuremath{-}}
\DeclareUnicodeCharacter{FB01}{fi}
\usepackage{amsmath}
\usepackage{amssymb}
\usepackage{graphicx}
\usepackage{xcolor}
\usepackage{hyperref}
\hypersetup{
  colorlinks=true,
  citecolor=blue,
  linkcolor=blue,
  urlcolor=blue,
  pdftitle={Nonreciprocity Propels Nematic Defects},
  pdfauthor={Yunpeng Cao, Yu Du, and Yao Li}
}
\makeatletter

\usepackage{soul}
\begin{document}
\title{Nonreciprocity Propels Nematic Defects}
\date{\today}

\author{Yunpeng Cao}
\thanks{These authors contributed equally to this work.}
\affiliation{School of Physics and Key Laboratory of Functional Polymer Materials
of Ministry of Education, Nankai University, and Collaborative Innovation
Center of Chemical Science and Engineering, Tianjin 300071, China}

\author{Yu Du}
\thanks{These authors contributed equally to this work.}
\affiliation{School of Physics and Key Laboratory of Functional Polymer Materials
of Ministry of Education, Nankai University, and Collaborative Innovation
Center of Chemical Science and Engineering, Tianjin 300071, China}
\affiliation{Jiangsu Key Laboratory for Biomaterials and Devices, School of Biological Science
and Medical Engineering, Southeast University, Nanjing, China}

\author{Yao Li}
\email[Contact author: ]{liyao@nankai.edu.cn}
\affiliation{School of Physics and Key Laboratory of Functional Polymer Materials
of Ministry of Education, Nankai University, and Collaborative Innovation
Center of Chemical Science and Engineering, Tianjin 300071, China}

\begin{abstract}
Nonreciprocal interactions are a generic source of nonequilibrium dynamics, yet their consequences for nematic topological defects remain unexplored. We introduce a nonreciprocal nematic model on a two-dimensional hexagonal lattice and formulate a continuum theory containing a nematic nonreciprocal torque. In the deterministic limit, simulations and symmetry analysis show that this torque drives +1/2 defects to self-propel along their polar axes. Nonreciprocity modifies the splitting pathway of integer +1 defects. These results identify a symmetry-resolved route from nonreciprocity to nematic defect transport and topological reorganization, distinct from active-stress-driven motion.
\end{abstract}
\maketitle

Nonreciprocal interactions, in which the influence exerted by one component on another differs from the reverse influence, provide a general route to nonequilibrium dynamics \cite{fruchart_non-reciprocal_2021,you_nonreciprocity_2020,saha_scalar_2020}. They occur in chemically interacting active colloids \cite{soto_self-assembly_2014}, programmable robotic mixtures \cite{chen_emergent_2024}, active-particle systems with vision-dependent interactions \cite{lavergne_group_2019,barberis_large-scale_2016}, as well as in ecological systems \cite{blumenthal_phase_2024}, asymmetric social-influence networks \cite{sornette_non-normal_2023}, and effective spin models \cite{loos_long-range_2023,avni_nonreciprocal_2025}. Such interaction asymmetry can break detailed balance and generate traveling states, non-Hermitian phase transitions, and nonequilibrium pattern formation \cite{fruchart_non-reciprocal_2021,you_nonreciprocity_2020,saha_scalar_2020}.

The consequences of nonreciprocity for topological defects have
recently begun to attract attention. In nonreciprocal XY models,
recent work has revealed directional defect propagation
\cite{loos_long-range_2023}, nonreciprocal selection of defect
textures and altered pair-annihilation dynamics
\cite{rouzaire_nonreciprocal_2025}, and defect cloaking that can
delay annihilation \cite{popli_ordering_2025,dopierala_inescapable_2025}. Extensions of
nonreciprocal XY models to motile lattice-gas systems have further
revealed stationary aster clusters with radial spin order under
vision-cone coupling \cite{ref18b}. In nonreciprocal polar flocks,
the breakdown of chiral order is accompanied by defect proliferation
and spatiotemporal chaos \cite{myin_breakdown_2026,woo_extensive_2026}.
Nematic order presents a
distinct setting because its head--tail symmetry permits
half-integer defects, in contrast to the integer defects of
XY-like order \cite{doostmohammadi_active_2018}. How interaction-level nonreciprocity affects these
nematic defects remains comparatively unexplored.

In conventional active nematics, active-stress-driven flows lead to the characteristic self-propulsion of $+1/2$ defects, whereas $-1/2$ defects do not exhibit comparable intrinsic self-propulsion \cite{tang_theory_2019,giomi_defect_2013,doostmohammadi_active_2018,giomi_defect_2014}. This contrast reflects their distinct local symmetries: a $+1/2$ defect possesses a polar axis, whereas a $-1/2$ defect has threefold symmetry and no unique polar direction \cite{vromans_orientational_2016,tang_orientation_2017}. Against this background, our central question is how interaction-level nonreciprocity couples to the distinct symmetries of nematic defects and thereby shapes their nonequilibrium dynamics.

In this work, we introduce a nonreciprocal nematic model on a hexagonal lattice, with soft vision-cone interactions \cite{rouzaire_nonreciprocal_2025} adapted to nematic symmetry. A long-wavelength expansion of the lattice dynamics yields the corresponding smooth-director angular equation, and we construct a symmetry-compatible regular $Q$-tensor formulation to resolve defect cores and dynamics. We find that nonreciprocity drives intrinsic polar motion of $+1/2$ defects while producing no corresponding intrinsic translation of isolated $-1/2$ defects. It further reshapes the splitting pathways of integer $+1$ defects. Together, these results connect nonreciprocity to symmetry-selective defect transport and multidefect topological reorganization.

\begin{figure}[tbp]
\centering
\includegraphics[width=\columnwidth]{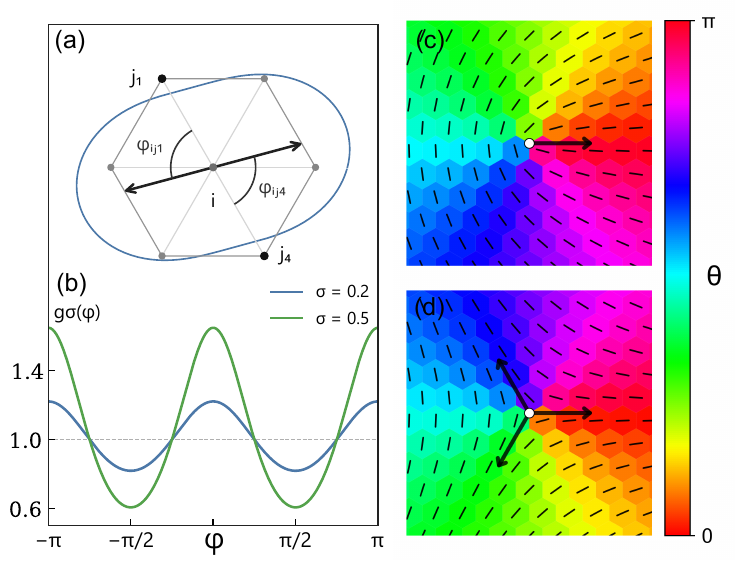}
\caption{Model and elementary defect textures.
(a) Hexagonal Bravais lattice geometry and the bond angle $\varphi_{ij}$ entering Eq.~\eqref{eq:lattice_dynamics}.
(b) Vision-cone interaction kernel $g_\sigma(\varphi)$ for $\sigma=0.2$ and $0.5$. Larger $\sigma$ concentrates the interaction weight more strongly along the local director direction.
(c) Representative $+1/2$ defect texture, which carries a well-defined polar axis set by the phase parameter $\mu$.
(d) Representative $-1/2$ defect texture, which has threefold symmetry and no unique polar axis.}
\label{fig:fig1}
\end{figure}

We consider a two-dimensional nematic order-parameter field on a hexagonal Bravais lattice, equivalently a triangular lattice, with six nearest neighbors per bulk site. Each lattice site $i$ carries a $\pi$-periodic director angle $\theta_i\in(-\pi/2,\pi/2]$, whose dynamics is governed by
\begin{equation}
\gamma\,\dot\theta_i =
J\sum_{j\in\partial_i} g_\sigma(\varphi_{ij})\,
\sin\!\left[2(\theta_j-\theta_i)\right]
+ \sqrt{2\gamma T}\,\eta_i(t),
\label{eq:lattice_dynamics}
\end{equation}
where $\partial_i$ denotes the six nearest neighbors of site $i$, $\varphi_{ij}$ is the bond angle from $i$ to $j$ measured relative to the local director at $i$, $\eta_i(t)$ is a unit-strength Gaussian white noise, $\gamma$ is the rotational friction coefficient, and $T$ is the noise temperature. Finite-temperature robustness is not addressed in the present work. We choose a vision-cone-type nonreciprocal interaction kernel~\cite{rouzaire_nonreciprocal_2025},
\begin{equation}
g_\sigma(\varphi_{ij})=\exp\!\left(\sigma\cos 2\varphi_{ij}\right),
\label{eq:interaction_kernel}
\end{equation}
which is $\pi$-periodic and therefore compatible with nematic symmetry. For $\sigma=0$, Eq.~\eqref{eq:lattice_dynamics} reduces to the passive nematic interaction. For $\sigma>0$, the coupling depends explicitly on the orientation of the focal site $i$: neighbors located along the local director are weighted more strongly than those in the transverse direction. As a result, the coupling from $i$ to $j$ generally differs from that from $j$ to $i$ [Figs.~\ref{fig:fig1}(a,b)].

Half-integer defects are identified from the nematic winding around each elementary triangular plaquette: a total winding of $\pm\pi$ corresponds to a topological charge of $\pm1/2$, with the defect position assigned to the plaquette centroid.

Direct simulations of Eq.~\eqref{eq:lattice_dynamics} show that an isolated 
$+1/2$ defect develops persistent motion along its polar axis (see Video S1 \cite{supp2026}). 
To examine the mechanism, we derive the long-wavelength angular dynamics outside defect cores and construct a minimal symmetry-compatible $Q$-tensor closure that regularizes the cores; details are given in Sec.~S1 of the Supplemental Material~\cite{supp2026}
\begin{equation}
\partial_t Q =
K\,\nabla^2 Q
+ \tilde\sigma\,\big[\nabla\times(\nabla\cdot Q)\big]\,Q^\perp
+ \alpha\,[1-2\,{\rm Tr}(\,Q^2)]\,Q,
\label{eq:q_tensor_dynamics}
\end{equation}
Here $K$ is the nematic elastic constant, $\alpha$ controls 
relaxation of the order-parameter magnitude, and $Q^\perp=\partial Q/\partial\theta$ at 
fixed magnitude is the rotational tangent, with $(Q^\perp)_{xx}=-2Q_{xy}$ and $(Q^\perp)_{xy}=2Q_{xx}$.  
In continuum units with $K=1$, we use $\tilde\sigma=\sigma/2$. Unless otherwise stated, reported nonreciprocal strengths refer to the microscopic parameter $\sigma$ in Eq.~\eqref{eq:interaction_kernel}.

\begin{figure}[tbp]
\centering
\includegraphics[width=\columnwidth]{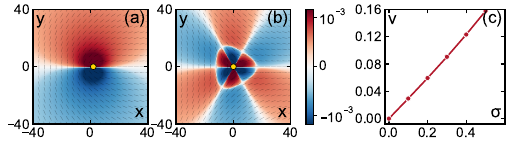}
\caption{Symmetry origin of $+1/2$ defect self-propulsion in the nonreciprocal nematic continuum theory [Eq.~\eqref{eq:q_tensor_dynamics}]. (a,b) Heat maps of $\partial_t\theta$ near isolated $+1/2$ and $-1/2$ defect cores, respectively, at $\sigma=0.5$. The $+1/2$ defect exhibits a polar dipolar torque imbalance, whereas the $-1/2$ defect exhibits a $D_3$-symmetric three-lobed pattern. (c) Fitted speed $v(\sigma)$ of an isolated $+1/2$ defect on a $1500\times750$ grid in $[-100,100)\times[-50,50)$, using the common post-relaxation, pre-boundary window $45.5\leq t\leq75.0$ for $\sigma\in[0.10,0.50]$. The speed increases with nonreciprocal strength.}
\label{fig:fig2}
\end{figure}

To identify which defect symmetries permit this motion, we examine isolated continuum defects under open boundary conditions. The initial director texture is $\theta_s=s\phi+\mu$, where $s=\pm1/2$ is the charge, $\phi$ is the polar angle about the core, and $\mu$ sets the texture orientation. The nonreciprocal term in Eq.~\eqref{eq:q_tensor_dynamics} generates a local orientational torque on the director field. In the overdamped description, the associated director-rotation rate is
\begin{align}
\dot\theta^{\rm nr}(\mathbf r)
&= \tilde\sigma\,\omega(\mathbf r),\notag\\
\omega
&= (\partial_x^2-\partial_y^2)Q_{xy}
-2\,\partial_{xy}Q_{xx}.
\label{eq:local_angular_velocity}
\end{align}
The derivation of Eq.~\eqref{eq:local_angular_velocity} is given in Sec.~S1 of the Supplemental Material.

A nonzero value of this local torque, which produces $\dot\theta^{\rm nr}\neq0$ in the overdamped dynamics, does not by itself imply motion of the defect core. Translation requires coordinated director rotations that reproduce the texture displaced to a new position. We identify this contribution by projecting the torque-induced field change onto the change associated with a small displacement, treating the defect position as a collective coordinate. For either charge, the angular part of this translational pattern is dipolar: it contains only $\sin\phi$ and $\cos\phi$, or harmonics $|m|=1$. Only the matching component of the angular-velocity field can therefore drive translation at leading order; the remaining components are orthogonal to it (Supplemental Material, Sec.~S2.1).

For a $+1/2$ defect, the nonreciprocal angular velocity has precisely this dipolar structure [Fig.~\ref{fig:fig2}(a)]. It couples to translation along the polar axis $\mathbf p=(\cos2\mu,\sin2\mu)$, allowing an intrinsic velocity $\mathbf V_{+1/2}=v(\sigma)\mathbf p$. The resulting self-propulsion is illustrated in Video S2 \cite{supp2026}, obtained from simulations of Eq.~\eqref{eq:q_tensor_dynamics}. For a $-1/2$ defect, by contrast, the torque has a threefold angular structure, $|m|=3$ [Fig.~\ref{fig:fig2}(b)]. Its three equivalent directions cancel in the translational projection: local rotations remain possible, but an isolated texture preserving this symmetry has no intrinsic self-propulsion. This is the point-group symmetry selection rule underlying the dynamical contrast \cite{ref2b,vromans_orientational_2016,tang_orientation_2017}. The derivation of the angular selection rule is given in Sec.~S2.1 of the Supplemental Material.

The simulations confirm the allowed positive-defect motion: its fitted speed increases monotonically with $\sigma$ [Fig.~\ref{fig:fig2}(c); measurement protocol in Supplemental Material, Sec.~S2.2]. Symmetry selects the axis of motion, whereas the dynamics determines the sign and magnitude of $v$. No hydrodynamic velocity field or prescribed active stress is introduced here; the positive defect moves through local orientational relaxation driven by the nonreciprocal torque, rather than advection by an active-stress-generated flow.

The preceding analysis focused on isolated defects. We now consider the minimal topologically neutral unit formed by a \(+1/2\) defect and a \(-1/2\) defect, to illustrate the geometry of its orientational texture. We initialize the pair through the orientational field
\begin{equation}
\begin{aligned}
\theta_0(\mathbf r)={}&\frac{1}{2}\operatorname{atan2}(y-y_+,x-x_+)\\
&-\frac{1}{2}\operatorname{atan2}(y-y_-,x-x_-)+\mu .
\end{aligned}
\label{eq:pair_initial_field}
\end{equation}
where \(\mathbf r_\pm=(x_\pm,y_\pm)\) denote the positions of the \(+1/2\) and \(-1/2\) defects, respectively, and \(\mu\) sets the overall orientational offset.

\begin{figure}[tbp]
\centering
\includegraphics[width=0.45\textwidth]{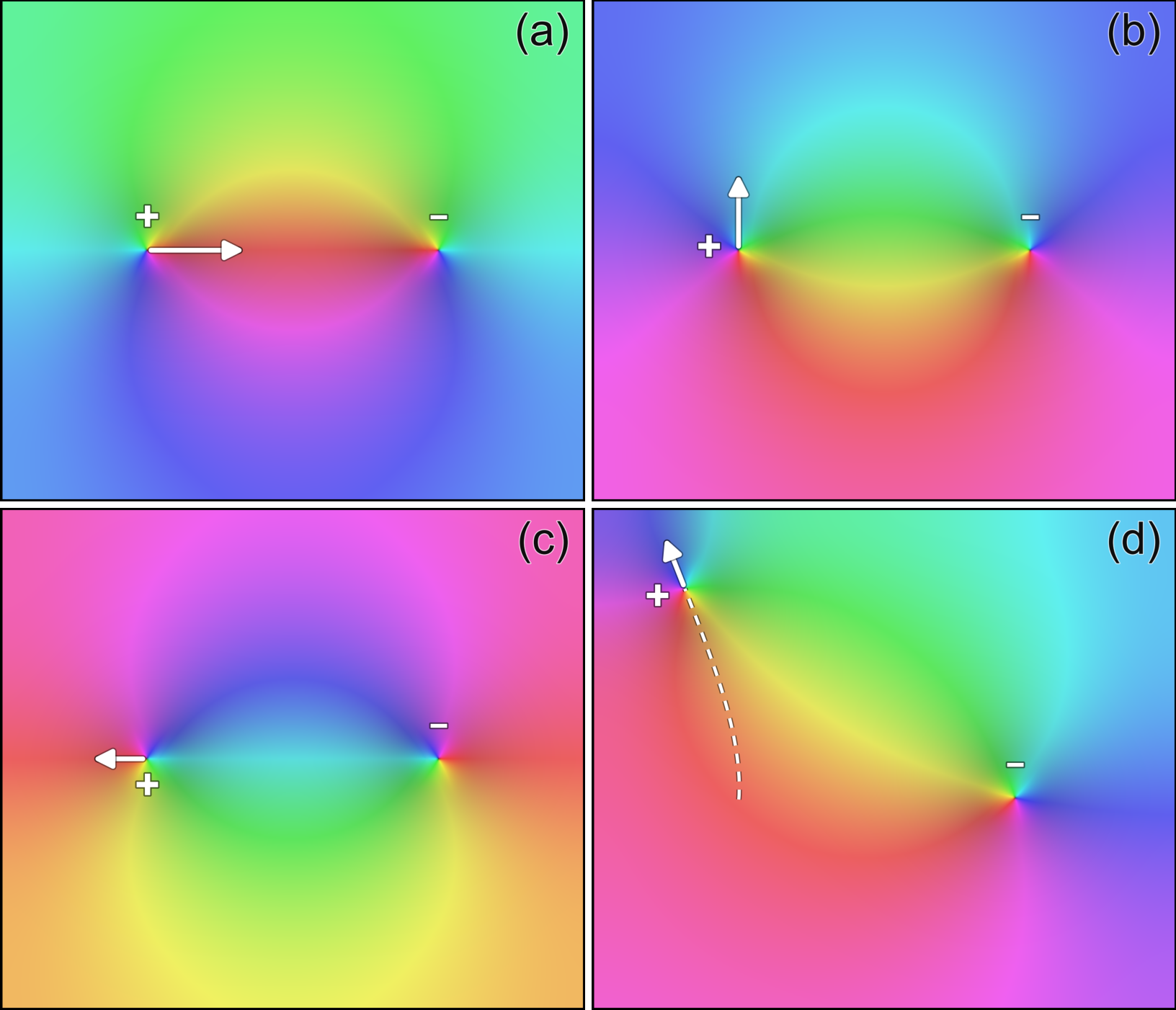}
\caption{Initial textures and an illustrative late-time reconstruction of a $\pm1/2$ defect pair.
(a--c) Initial textures at $t=0$ with local $+1/2$ polarity parallel ($\beta=0$), perpendicular ($\beta=\pi/2$), and antiparallel ($\beta=\pi$) to the initial pair axis, directed from the positive to the negative defect.
(d) Illustrative late-time reconstruction corresponding to (b) at $t=250$, with a displaced positive core and a rotated local polarity. Hue encodes the director angle $\theta\bmod\pi$ using a common cyclic color map. White circles mark defect cores, and white arrows indicate the local polarity $\mathbf p_+$ rather than the velocity. The dashed white line in (d) traces the positive-defect trajectory from $t=0$ to $t=250$ (see Supplemental Video S3 \cite{supp2026}). All panels show the same spatial window, $[-50,50]^2$. The continuum run for (d) uses $\sigma=0.50$, $N=1000$, $K=1$, $\alpha=10$, $r_0=50$, $T=0$, and $\Delta t=10^{-3}$.}
\label{fig:fig3}
\end{figure}

We define $\beta$ as the angle between the local polarity of the $+1/2$ defect and the line directed from the positive to the negative defect. Figure~\ref{fig:fig3}(a--c) shows the initial configurations generated from Eq.~\eqref{eq:pair_initial_field} for $\beta=0$, $\pi/2$, and $\pi$, respectively.

\begin{figure*}[tbp]
\centering
\includegraphics[width=\textwidth]{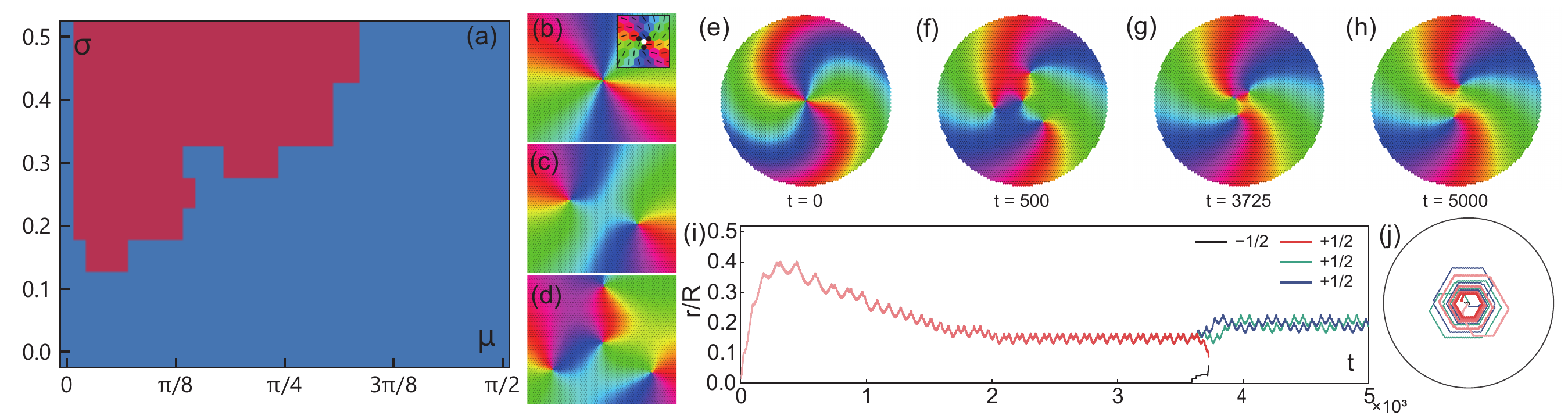}
\caption{Phase structure and boundary-mediated defect evolution.
(a) Splitting phase diagram in the $\mu$--$\sigma$ plane. Blue denotes the standard binary splitting mode, $+1\to2\times(+1/2)$, whereas orange-red denotes the four-defect mode, $+1\to-1/2+3\times(+1/2)$. The scan uses 33 uniformly spaced values of $\mu\in[0,\pi/2]$ and 11 values of $\sigma\in[0.00,0.50]$; simulations are performed with $L=100$, $T=0$, open boundaries implemented by excluding the outermost lattice sites, and evolved for $5\times10^3$ deterministic time steps for each parameter set.
(b--d) Representative director-field snapshots of the initial $+1$ configuration, the standard binary splitting mode, and the four-defect mode at high $\sigma$ (Supplemental Video S4--S5 \cite{supp2026}). 
(e--h) Representative snapshots of the evolution of the four-defect configuration under circular tangential anchoring (Supplemental Video S7 \cite{supp2026}), showing outward separation (e), elastic return (f), defect annihilation (g), and the subsequent two-$+1/2$ orbital motion (h).
(i) Radial defect distance $r/R$ as a function of time, where $R$ is the confinement radius.
(j) Tracked trajectories of the initial three $+1/2$ defects and one $-1/2$ defect, including the late-time orbital motion of the two surviving $+1/2$ defects near the boundary. For panels (e--j), $L=100$, $R=L/2.5=40$, and the system is observed up to $t=5000$.}
\label{fig:fig4}
\end{figure*}

These initial geometries provide a qualitative picture of the competition between polar driving and elastic attraction. If the positive defect retains a drive along its local polarity, parallel polarity is expected to assist the elastic approach, whereas antiparallel polarity is expected to oppose it. For perpendicular polarity, the polar drive initially points transverse to the pair axis, although elastic attraction and the coupled pair texture can still produce axial motion. These expectations concern the direction of the polar contribution rather than a quantitative law for the total pair velocity.

For the initially perpendicular pair, Figs.~\ref{fig:fig3}(b, d) and Video S3 \cite{supp2026} show the $+1/2$ defect displaced mainly transverse to the initial pair axis. The local polarity of the positive defect also turns away from its initial direction. Panel (d) is reconstructed from the saved core positions and local polarity rather than a stored field snapshot. The physical origin of the deflection is the modulation of the positive defect's local polarity by the nonuniform background orientational field. Through elastic coupling, the field generated by the companion defect and the surrounding texture rotates the local polarity as the pair configuration evolves. The nonreciprocal drive along that polarity consequently changes direction. Thus, the initial polarity sets the initial driving direction but does not require rectilinear defect motion at later times, consistent with the more general occurrence of curved and symmetry-broken defect trajectories in active systems \cite{beer_irreversibility_2026}. The snapshots do not quantitatively separate the contributions of the companion defect and the boundary to the background field.

Building on the isolated-defect results, we next examine the splitting of an initial nematic defect configuration with total topological charge $+1$ on the hexagonal lattice. The initial director field is chosen as
$
\theta_0(\mathbf{r}) = \operatorname{atan2}(y,x) + \mu ,
$
where the global phase $\mu$ controls the initial texture of the $+1$ defect; $\mu=0,\pi/4,\pi/2$ correspond to aster, spiral, and vortex configurations, respectively. 

Figure~\ref{fig:fig4} summarizes the splitting phase structure in the $\mu$--$\sigma$ parameter plane and further illustrates the subsequent evolution of the four-defect configuration under circular confinement.

In the weakly nonreciprocal regime, the system primarily follows the standard binary splitting mode,
$
+1 \rightarrow 2\times(+1/2)
$,
that is, one $+1$ defect splits into two $+1/2$ defects [Figs.~\ref{fig:fig4}(b,c) and Video S4 \cite{supp2026}]. This is consistent with the energetically favorable binary splitting reported in passive nematics \cite{kralj_decomposition_2017}.

As the nonreciprocal coupling is increased to $\sigma\gtrsim0.15$, a second splitting mode emerges:
$
+1 \rightarrow -1/2 + 3\times(+1/2)
$.
This configuration consists of one $-1/2$ defect near the center and three $+1/2$ defects 
separating along directions approximately $120^\circ$ apart, forming a threefold-symmetric outgoing configuration [Fig.~\ref{fig:fig4}(d) and Video S5 \cite{supp2026}]. Its total topological charge satisfies
$3\times(+1/2)+(-1/2)=+1$.
The triangular Bravais lattice has sixfold rotational symmetry, and the observed threefold outgoing configuration is compatible with its $C_3$ rotational subgroup. Once three $+1/2$ defects form, charge conservation requires the remaining defects to carry a net charge of $-1/2$, here carried by a single defect near the center. We refer to this pathway as the $3(+1/2)-1/2$ four-defect mode, or simply the four-defect mode, to distinguish it from the standard binary splitting mode.

Figure~\ref{fig:fig4}(a) shows the distribution of the two splitting modes in the 
$\mu$--$\sigma$ plane. The fraction of sampled $\mu$ values yielding the four-defect mode 
increases with $\sigma$: it is not observed at $\sigma=0.10$, whereas at $\sigma=0.50$ it accounts for $64\%$ of the sampled $\mu$ values. Correspondingly, $\sigma\in[0.15,0.35]$ forms the main crossover region between the two splitting modes. These results show that nonreciprocal coupling not only modifies the motion of defects after they form, but can also alter the topological reorganization pathway of an integer defect.

This mode selection can be understood from the nonreciprocal self-propulsion of $+1/2$ defects. In the standard binary splitting mode, the system lowers its elastic energy by producing two $+1/2$ defects. Under stronger nonreciprocal coupling, however, the newly formed $+1/2$ defects also carry an effective drive along their polar axes and therefore tend to move outward from the original core region. The three positive defects separate into a threefold arrangement compatible with the rotational symmetry allowed by the triangular lattice. This directional organization supports a lattice-assisted interpretation of the splitting pathway, but the dynamical selection of the three-defect outgoing channel cannot be determined from substrate symmetry and charge conservation alone. The arrangement does not by itself establish a generic continuum instability. 
The global phase $\mu$ determines the orientational organization within the initial $+1$ texture and therefore affects whether the four-defect mode can be activated. In particular, textures close to the spiral configuration ($\mu\sim\pi/4$) more readily support directed outward motion, whereas textures close to the vortex configuration ($\mu\to\pi/2$) are dominated by azimuthal organization and tend to remain in the standard binary splitting mode over the parameter range studied here.

The four-defect configuration also exhibits confinement size dependent evolution. 
To examine the role of the boundary, we introduce a circular domain anchoring, 
where the director at the boundary is perpendicular to the radial direction. Such confined 
geometries can organize defect motion and circulation in active-nematic systems \cite{schimming_analytical_2025,keber_topology_2014,duclos_spontaneous_2018,opathalage_self-organized_2019}. 
For the confined system shown in Figs.~\ref{fig:fig4}(e--j) and Video S7 \cite{supp2026}, the initial four-defect 
configuration expands as the three $+1/2$ defects move outward. Boundary confinement and elastic 
relaxation subsequently redirect the outward-moving defects toward the interior, as reflected in 
the decrease of their radial distances. The central $-1/2$ defect then forms a tight dipolar 
pair with one of the $+1/2$ defects and annihilates with it, leaving two $+1/2$ defects in 
persistent orbital motion near the boundary. We also repeated the confined simulation in a larger 
circular domain with $L=200$ and $R=80$, keeping all other parameters unchanged. Over the same 
observation interval up to $t=5000$, the four-defect configuration persists after splitting 
without annihilation (Supplemental Material, Sec.~S3 and Video S8 \cite{supp2026}). This comparison shows that the subsequent 
evolution of the splitting products depends on the confinement size over the range examined.

\textit{Conclusions---}Our results show that nonreciprocity propels $+1/2$ nematic defects and reshapes defect dynamics across multiple levels. This polarity-selective driving further modifies defect-pair dynamics, the splitting of integer $+1$ defects, and the confinement-dependent evolution of the resulting multi-defect configurations. Together, these results establish nonreciprocity as a distinct mechanism for generating and regulating defect motion.

This mechanism differs fundamentally from that of conventional active nematics, 
in which active stresses, $\sigma_{ij}^{\rm a}=-\zeta Q_{ij}$, generate flows and backflows that advect 
defect textures \cite{aditi_simha_hydrodynamic_2002,giomi_defect_2013,thampi_velocity_2013,doostmohammadi_stabilization_2016}. 
In the present framework, by contrast, neither a hydrodynamic velocity field nor an explicit active stress is introduced. 
Instead, the directed motion of the $+1/2$ defect emerges directly from a symmetry-selected nonreciprocal torque. 
Our results therefore establish a direct route from interaction nonreciprocity to defect self-propulsion.

More broadly, our findings show that the interplay between nonreciprocity and defect symmetry provide a theoretical framework for understanding and organizing topological dynamics in nonequilibrium systems, and also offer a design principle for engineering defect transport and organization in bioinspired active-nematic systems.

\vspace{0.5\baselineskip}
\textit{Acknowledgments---}This work was financially supported by the National Natural Science Foundation of China (12275137). We thank Jeff Z. Y. Chen, Baohui Li and Zheng Wang for technical help.

\makeatletter
\immediate\write\@auxout{\string\citation{apsrev42Control}}
\makeatother
\bibliographystyle{apsrev4-2}
\bibliography{reference}

@article{fruchart_non-reciprocal_2021,
  author  = {Fruchart, Michel and Hanai, Ryo and Littlewood, Peter B. and Vitelli, Vincenzo},
  title   = {Non-reciprocal phase transitions},
  journal = {Nature},
  volume  = {592},
  pages   = {363--369},
  year    = {2021},
  doi     = {10.1038/s41586-021-03375-9},
}

@article{you_nonreciprocity_2020,
  author  = {You, Zhihong and Baskaran, Aparna and Marchetti, M. Cristina},
  title   = {Nonreciprocity as a generic route to traveling states},
  journal = {Proc. Natl. Acad. Sci. U.S.A.},
  volume  = {117},
  pages   = {19767--19772},
  year    = {2020},
  doi     = {10.1073/pnas.2010318117},
}

@article{saha_scalar_2020,
  author  = {Saha, Suropriya and Agudo-Canalejo, Jaime and Golestanian, Ramin},
  title   = {Scalar {Active} {Mixtures}: {The} {Nonreciprocal} {Cahn--Hilliard} {Model}},
  journal = {Phys. Rev. X},
  volume  = {10},
  pages   = {041009},
  year    = {2020},
  doi     = {10.1103/PhysRevX.10.041009},
}

@article{soto_self-assembly_2014,
  author  = {Soto, Rodrigo and Golestanian, Ramin},
  title   = {Self-Assembly of Catalytically Active Colloidal Molecules: Tailoring Activity Through Surface Chemistry},
  journal = {Phys. Rev. Lett.},
  volume  = {112},
  pages   = {068301},
  year    = {2014},
  doi     = {10.1103/PhysRevLett.112.068301},
}

@article{chen_emergent_2024,
  author  = {Chen, Jianchao and Lei, Xiaokang and Xiang, Yalun and Duan, Mengyuan and Peng, Xingguang and Zhang, H. P.},
  title   = {Emergent Chirality and Hyperuniformity in an Active Mixture with Nonreciprocal Interactions},
  journal = {Phys. Rev. Lett.},
  volume  = {132},
  pages   = {118301},
  year    = {2024},
  doi     = {10.1103/PhysRevLett.132.118301},
}

@article{lavergne_group_2019,
  author  = {Lavergne, François A. and Wendehenne, Hugo and Bäuerle, Tobias and Bechinger, Clemens},
  title   = {Group formation and cohesion of active particles with visual perception-dependent motility},
  journal = {Science},
  volume  = {364},
  pages   = {70--74},
  year    = {2019},
  doi     = {10.1126/science.aau5347},
}

@article{barberis_large-scale_2016,
  author  = {Barberis, Lucas and Peruani, Fernando},
  title   = {Large-Scale Patterns in a Minimal Cognitive Flocking Model: Incidental Leaders, Nematic Patterns, and Aggregates},
  journal = {Phys. Rev. Lett.},
  volume  = {117},
  pages   = {248001},
  year    = {2016},
  doi     = {10.1103/PhysRevLett.117.248001},
}

@article{blumenthal_phase_2024,
  author  = {Blumenthal, Emmy and Rocks, Jason W. and Mehta, Pankaj},
  title   = {Phase Transition to Chaos in Complex Ecosystems with Nonreciprocal Species--Resource Interactions},
  journal = {Phys. Rev. Lett.},
  volume  = {132},
  pages   = {127401},
  year    = {2024},
  doi     = {10.1103/PhysRevLett.132.127401},
}

@article{sornette_non-normal_2023,
  author  = {Sornette, Didier and Lera, Sandro Claudio and Lin, Jianhong and Wu, Ke},
  title   = {Non-normal interactions create socio-economic bubbles},
  journal = {Commun. Phys.},
  volume  = {6},
  pages   = {261},
  year    = {2023},
  doi     = {10.1038/s42005-023-01379-7},
}

@article{loos_long-range_2023,
  author  = {Loos, Sarah A. M. and Klapp, Sabine H. L. and Martynec, Thomas},
  title   = {Long-Range Order and Directional Defect Propagation in the Nonreciprocal {XY} Model with Vision Cone Interactions},
  journal = {Phys. Rev. Lett.},
  volume  = {130},
  pages   = {198301},
  year    = {2023},
  doi     = {10.1103/PhysRevLett.130.198301},
}

@article{avni_nonreciprocal_2025,
  author  = {Avni, Yael and Fruchart, Michel and Martin, David and Seara, Daniel and Vitelli, Vincenzo},
  title   = {Nonreciprocal {Ising} Model},
  journal = {Phys. Rev. Lett.},
  volume  = {134},
  pages   = {117103},
  year    = {2025},
  doi     = {10.1103/PhysRevLett.134.117103},
}

@article{rouzaire_nonreciprocal_2025,
  author  = {Rouzaire, Ylann and Pearce, Daniel J. G. and Pagonabarraga, Ignacio and Levis, Demian},
  title   = {Nonreciprocal Interactions Reshape Topological Defect Annihilation},
  journal = {Phys. Rev. Lett.},
  volume  = {134},
  pages   = {167101},
  year    = {2025},
  doi     = {10.1103/PhysRevLett.134.167101},
}

@article{popli_ordering_2025,
  author  = {Popli, Pankaj and Maitra, Ananyo and Ramaswamy, Sriram},
  title   = {Ordering and Defect Cloaking in Nonreciprocal Lattice {XY} Models},
  journal = {Phys. Rev. Lett.},
  volume  = {135},
  pages   = {088303},
  year    = {2025},
  doi     = {10.1103/2yky-45sr},
}

@article{ref18b,
  author  = {Du, Yu and Cao, Yunpeng and Liu, Wenyu and Li, Yao},
  title   = {Collective Behavior of the Non-Reciprocal Vision Cone {XY} Lattice-Gas Model},
  journal = {Newton},
  volume  = {2},
  pages   = {100291},
  year    = {2026},
  doi     = {10.1016/j.newton.2025.100291},
}

@article{myin_breakdown_2026,
  author  = {Myin, Charlotte and Saha, Suropriya and Mahault, Benoît},
  title   = {Breakdown of Emergent Chiral Order and Defect Chaos in Nonreciprocal Flocks},
  journal = {Phys. Rev. Lett.},
  volume  = {137},
  pages   = {118302},
  year    = {2026},
  doi     = {10.1103/c6dy-38z1},
}

@article{woo_extensive_2026,
  author  = {Woo, Chul-Ung and Noh, Jae Dong and Rieger, Heiko},
  title   = {Extensive Spatiotemporal Chaos in Nonreciprocal Flocking},
  journal = {Phys. Rev. Lett.},
  volume  = {137},
  pages   = {118301},
  year    = {2026},
  doi     = {10.1103/lfwy-bbmv},
}

@article{doostmohammadi_active_2018,
  author  = {Doostmohammadi, Amin and Ignés-Mullol, Jordi and Yeomans, Julia M. and Sagués, Francesc},
  title   = {Active nematics},
  journal = {Nat. Commun.},
  volume  = {9},
  pages   = {3246},
  year    = {2018},
  doi     = {10.1038/s41467-018-05666-8},
}

@article{tang_theory_2019,
  author  = {Tang, Xingzhou and Selinger, Jonathan V.},
  title   = {Theory of defect motion in {2D} passive and active nematic liquid crystals},
  journal = {Soft Matter},
  volume  = {15},
  pages   = {587--601},
  year    = {2019},
  doi     = {10.1039/C8SM01901K},
}

@article{giomi_defect_2013,
  author  = {Giomi, Luca and Bowick, Mark J. and Ma, Xu and Marchetti, M. Cristina},
  title   = {Defect Annihilation and Proliferation in Active Nematics},
  journal = {Phys. Rev. Lett.},
  volume  = {110},
  pages   = {228101},
  year    = {2013},
  doi     = {10.1103/PhysRevLett.110.228101},
}

@article{giomi_defect_2014,
  author  = {Giomi, Luca and Bowick, Mark J. and Mishra, Prashant and Sknepnek, Rastko and Marchetti, M. Cristina},
  title   = {Defect dynamics in active nematics},
  journal = {Philos. Trans. R. Soc. A},
  volume  = {372},
  pages   = {20130365},
  year    = {2014},
  doi     = {10.1098/rsta.2013.0365},
}

@article{vromans_orientational_2016,
  author  = {Vromans, Arthur J. and Giomi, Luca},
  title   = {Orientational properties of nematic disclinations},
  journal = {Soft Matter},
  volume  = {12},
  pages   = {6490--6495},
  year    = {2016},
  doi     = {10.1039/C6SM01146B},
}

@article{tang_orientation_2017,
  author  = {Tang, Xingzhou and Selinger, Jonathan V.},
  title   = {Orientation of topological defects in {2D} nematic liquid crystals},
  journal = {Soft Matter},
  volume  = {13},
  pages   = {5481--5490},
  year    = {2017},
  doi     = {10.1039/C7SM01195D},
}

@article{ref2b,
  author  = {Kleman, M. and Lavrentovich, O. D.},
  title   = {Topological point defects in nematic liquid crystals},
  journal = {Philos. Mag.},
  volume  = {86},
  pages   = {4117--4137},
  year    = {2006},
  doi     = {10.1080/14786430600593016},
}

@article{beer_irreversibility_2026,
  author       = {Be'er, A. and Neimand, E. D. and Agarwal, Y. and Corbett, D. and Pearce, D. J. G. and Ariel, G. and Yashunsky, V.},
  title        = {Irreversibility and symmetry breaking in the creation and annihilation of defects in active living matter},
  journal      = {Nat. Phys.},
  year         = {2026},
  doi          = {10.1038/s41567-026-03378-1},
}

@article{kralj_decomposition_2017,
  author  = {Kralj, Samo and Murray, Bryce S. and Rosenblatt, Charles},
  title   = {Decomposition of strongly charged topological defects},
  journal = {Phys. Rev. E},
  volume  = {95},
  pages   = {042702},
  year    = {2017},
  doi     = {10.1103/PhysRevE.95.042702},
}

@article{schimming_analytical_2025,
  author  = {Schimming, Cody D. and Reichhardt, C. J. O. and Reichhardt, C.},
  title   = {Analytical model for the motion and interaction of two-dimensional active nematic defects},
  journal = {Soft Matter},
  volume  = {21},
  pages   = {122--136},
  year    = {2025},
  doi     = {10.1039/D4SM00956H},
}

@article{keber_topology_2014,
  author  = {Keber, Felix C. and Loiseau, Etienne and Sanchez, Tim and DeCamp, Stephen J. and Giomi, Luca and Bowick, Mark J. and Marchetti, M. Cristina and Dogic, Zvonimir and Bausch, Andreas R.},
  title   = {Topology and dynamics of active nematic vesicles},
  journal = {Science},
  volume  = {345},
  pages   = {1135--1139},
  year    = {2014},
  doi     = {10.1126/science.1254784},
}

@article{duclos_spontaneous_2018,
  author  = {Duclos, G. and Blanch-Mercader, C. and Yashunsky, V. and Salbreux, G. and Joanny, J.-F. and Prost, J. and Silberzan, P.},
  title   = {Spontaneous shear flow in confined cellular nematics},
  journal = {Nat. Phys.},
  volume  = {14},
  pages   = {728--732},
  year    = {2018},
  doi     = {10.1038/s41567-018-0099-7},
}

@article{opathalage_self-organized_2019,
  author  = {Opathalage, Achini and Norton, Michael M. and Juniper, Michael P. N. and Langeslay, Blake and Aghvami, S. Ali and Fraden, Seth and Dogic, Zvonimir},
  title   = {Self-organized dynamics and the transition to turbulence of confined active nematics},
  journal = {Proc. Natl. Acad. Sci. U.S.A.},
  volume  = {116},
  pages   = {4788--4797},
  year    = {2019},
  doi     = {10.1073/pnas.1816733116},
}

@article{aditi_simha_hydrodynamic_2002,
  author  = {Simha, R. Aditi and Ramaswamy, Sriram},
  title   = {Hydrodynamic Fluctuations and Instabilities in Ordered Suspensions of Self-Propelled Particles},
  journal = {Phys. Rev. Lett.},
  volume  = {89},
  pages   = {058101},
  year    = {2002},
  doi     = {10.1103/PhysRevLett.89.058101},
}

@article{thampi_velocity_2013,
  author  = {Thampi, Sumesh P. and Golestanian, Ramin and Yeomans, Julia M.},
  title   = {Velocity Correlations in an Active Nematic},
  journal = {Phys. Rev. Lett.},
  volume  = {111},
  pages   = {118101},
  year    = {2013},
  doi     = {10.1103/PhysRevLett.111.118101},
}

@article{doostmohammadi_stabilization_2016,
  author  = {Doostmohammadi, Amin and Adamer, Michael F. and Thampi, Sumesh P. and Yeomans, Julia M.},
  title   = {Stabilization of active matter by flow-vortex lattices and defect ordering},
  journal = {Nat. Commun.},
  volume  = {7},
  pages   = {10557},
  year    = {2016},
  doi     = {10.1038/ncomms10557},
}

@article{dopierala_inescapable_2025,
  title   = {Inescapable anisotropy of nonreciprocal {XY} models},
  author  = {Dopierala, Dawid and Chat\'e, Hugues and Shi, Xia-qing and Solon, Alexandre},
  journal = {Phys. Rev. Lett.},
  volume  = {135},
  number  = {8},
  pages   = {088302},
  year    = {2025},
  doi     = {10.1103/r3dx-7lrd},
}

@misc{supp2026,
  title   = {See Supplemental Material: continuum derivation, defect-symmetry projection, speed-extraction protocol, splitting and confinement controls, and supplemental Videos S1--S8.},
}

@CONTROL{apsrev42Control,
  author = {08},
  editor = {1},
  pages  = {0},
  title  = {0},
  year   = {1},
}
\end{document}